\documentclass[aps,nofootinbib,twocolumn,prd,eqsecnum,showpacs,showkeys,preprintnumbers,nofootinbib]{revtex4-1}
\usepackage{graphicx}
\usepackage{amsmath}
\usepackage{amsfonts}
\usepackage{amssymb}
\usepackage{color}
\usepackage{mathrsfs}
\usepackage{epstopdf}
\usepackage{url}
\usepackage{footnote}
\usepackage{textcomp}
\usepackage{hyperref}
\usepackage{dcolumn}
\usepackage[caption=false]{subfig}
\usepackage{cases}
\usepackage{bm}

\begin{document}

\title{The energy spectrum of gravitational waves in a loop quantum cosmological
model}

\author{Jo\~{a}o Morais $^{1}$}
\email{joao.morais@ist.utl.pt}
\author{Mariam Bouhmadi-L\'{o}pez $^{2,3}$}
\email{mariam.bouhmadi@ehu.es}
\author{Alfredo B. Henriques $^{1}$}
\email{alfredo.henriques@fisica.ist.utl.pt}

\date{\today}

\affiliation{
${}^1$ Centro Multidisciplinar de Astrof\'{\i}sica - CENTRA, Departamento de F\'{\i}sica, Instituto Superior T\'ecnico, Av. Rovisco Pais 1, 1049-001 Lisboa, Portugal\\
${}^2$ Department of Theoretical Physics, University of the Basque Country
UPV/EHU, P.O. Box 644, 48080 Bilbao, Spain\\
${}^3$ IKERBASQUE, Basque Foundation for Science, 48011, Bilbao, Spain\\}

\begin{abstract}
We explore the consequences of loop quantum cosmology (inverse-volume corrections) in the spectrum of the gravitational waves using the method of the Bogoliubov coefficients. These corrections are taken into account at the background level of the theory as well as at the first order in the perturbations theory framework. We show that these corrections lead to an intense graviton production during the loop super-inflationary phase prior to the standard slow-roll era, which leave their imprints through new features on the energy spectrum of the gravitational waves as would be measured today, including a new maximum on the low frequency end of the spectrum.
\end{abstract}


\maketitle

%
%

\section{INTRODUCTION}
\label{Introduction}

Gravitational waves (GWs) are, at the present time, the subject of an important research effort \cite{Abbott:2009ws,Aasi:2013sia}. In the field of cosmology, they may provide us with important information on the very early stages after the big-bang, information that might be unobtainable by other means \cite{Sathyaprakash:2009xs,Liddle,Krauss:2013pha}. We also witness an increased interest in the application of the ideas of loop quantum gravity to the problems of cosmology, a field known as loop quantum cosmology (LQC), after a series of seminal papers by Martin Bojowald \cite{Bojowald:1999tr,Bojowald:2000pk,Bojowald:2001xe,Bojowald:2001xa,Bojowald:2001vw,Bojowald:2001ep,Bojowald:2002gz,Bojowald:2002nz}. For a review on LQC, see for example \cite{Ashtekar:2003hd,Ashtekar:2004eh,Bojowald:2008zzb,Banerjee:2011qu}. Among the important results given by LQC, we have the possibility of removing in a natural way the initial singularity \cite{Bojowald:2001xe,Ashtekar:2004eh,Ashtekar:2003hd,Bojowald:2003uh,Bojowald:2004ax,Ashtekar:2006wn}. LQC introduces, in the semi-classical period prior to the classical slow-roll inflation, important modifications in the dynamical equations driving the expansion of the universe, for example it induces a super-inflationary period \cite{Tsujikawa:2003vr}, and such changes, in turn, give rise to an extra production of GWs, even without the appropriate modifications into the gravitational equations, as has been shown in Refs.~\cite{Afonso:2010fa,Sa:2011rm}. The study of GWs in LQC has been a very active field in the past few years \cite{Grain:2009eg,Afonso:2010fa,Bojowald:2011hd,Mielczarek:2007zy,Mielczarek:2007wc,Grain:2009cj,Mielczarek:2008pf,Grain:2009kw,Mielczarek:2009vi,Mielczarek:2010bh,Linsefors:2012et,Cailleteau:2013kqa,Grain:2009kw,Grain:2010yv,Sa:2011rm} including the analysis of the power spectrum of the tensor modes (i) with inverse-volume corrections Refs.~\cite{Sa:2011rm,Mielczarek:2007zy,Mielczarek:2007wc,Grain:2009cj,Bojowald:2011hd}, (ii) with holonomy corrections Refs.~\cite{Mielczarek:2008pf,Grain:2009kw,Mielczarek:2009vi,Mielczarek:2010bh,Linsefors:2012et}, and (iii) considering both these corrections simultaneously \cite{Grain:2009eg}. More recently, the evolution equations for the tensorial perturbations including inverse-volume and holonomy corrections within a generalised anomaly-free formalism have been deduced in Ref.~\cite{Cailleteau:2013kqa}. The possible footprints of LQC on the B-modes polarization of the cosmic microwave background (CMB) has been tackled in Ref.~\cite{Grain:2010yv}.

In the present paper, we analyse the spectrum of the GWs as would be measured today. More precisely, we modify the equations for the GWs, introducing inverse-volume corrections (we leave to a future paper the holonomy corrections), and compare the results with those obtained in Refs.~\cite{Afonso:2010fa,Sa:2011rm}, where these corrections were introduced only in the background dynamical equations driving the expansion of the universe. What we see is an important extra production of GWs, leaving its imprint in the low-frequency limit of today's energy-spectrum, which, incidentally, also shows that inflation does not remove all the information coming from phenomena taking place in the pre-inflationary times. Indeed, as we have shown recently, a bounce in modified theories of gravity \cite{BouhmadiLopez:2012qp} as well as a topological defect phase prior to classical inflation \cite{BouhmadiLopez:2012by} leave some imprints on the low frequencies of the spectrum of the GWs which are not washed out by the inflationary phase. What happens is that the physical features during the semi-classical period affect in different ways different frequencies, and the memory of these differences survives through the inflationary period, to be shown today in the power-spectrum. Besides this extra production of gravitons, when compared with classical models, after the usual initial decrease in the energy-spectrum of the very low frequencies, we have then a second maximum, brought about by LQC, which is not present when we discard the modifications, brought in by loop quantum cosmology, in the GW equations. In Sec.~\ref{sec3}, we suggest an explanation for this interesting new feature. 

In our model, inflation is driven by a chaotic type of potential, of the
form $(1/2)m_{\phi }^{2}\phi ^{2},$ although we believe that the main
results will not be much modified by the use of a different potential.
Results from the Planck satellite collaboration \cite{Ade:2013uln} do not seem to
particularly favour this potential, but also do not disfavour it completely.
For this reason we keep it as a simple toy model, and also because most of
the analyses presented were made in the context of classical inflation \cite{Ijjas:2013vea},
which is not the context of the present paper.

We organize the paper as follows. The LQC model used in
our work is described in Sec.~\ref{sec2}, where we summarise the equations
of motion for both the semi-classical and the classical stages of the
evolution of the universe, and where the values of the parameters and the
intial conditions for the numerical integrations are specified, taking into account various cosmological observations like measurements of the CMB. In
Sec.~\ref{sec3} we review the evolution equations for the tensorial modes in LQC taken into account inverse-volume corrections. Then, we calculate their energy-spectrum, as
would be seen today, using the method of the continuous Bogoliubov coefficients,
first derived by Leonard Parker \cite{Parker:1969au}. We compare with previous results
obtained by one of us in \cite{Sa:2011rm} and comment on the differences obtained. Sec.~\ref{sec4}
summarizes the main results of the paper.

%
%

\section{\label{sec2}The Model}

To describe the early stages of the evolution of the universe, the equations
of standard cosmology have to be modified by corrections due to the loop
quantum effects, defining the semi-classical stage of the expansion \cite{Bojowald:2008zzb,Banerjee:2011qu}. After a few Planck times, we enter into the usual classical regime, with a period of inflation driven, in our paper, by a scalar field $\phi $ with a chaotic-type potential
\begin{equation}
	V(\phi )=\frac{1}{2}m_{\phi }^{2}\phi ^{2},  \label{1}
\end{equation}
followed by a period of reheating and, finally, by radiation-dominated,
matter-dominated and dark-energy-dominated periods.

As we said in the introduction, in the present paper we take into account
only the inverse-volume corrections for the initial semi-classical stage. The modified Friedmann and Raychaudhury
equations are then given by \cite{Bojowald:2008zzb,Banerjee:2011qu}
\begin{equation}
	\label{2}
	\left(\frac{\dot{a}}{a}\right)^2=\frac{8\pi }{3m_{P}^{2}}\left[\frac{\dot{\phi} ^{2}}{2d(q)}+V(\phi )\right],  
\end{equation}
\begin{equation}
	\label{3}
	\frac{\ddot{a}}{a}=\frac{8\pi }{3m_{P}^{2}}\left[V(\phi )-\frac{\phi ^{2}}{d(q)}\right]+\frac{2\pi \dot{\phi }^{2}}{m_{P}^{2}}\frac{f(q)}{d(q)},  
\end{equation}
while the evolution of the scalar field is dictated by the equation
\begin{equation}
	\label{4}
	\ddot{\phi }=-3\frac{\dot{a}}{a}\left[1-f(q)\right]\dot{\phi }-d(q)\frac{\partial V}{\partial \phi },  
\end{equation}
where we assumed a flat Friedmann-Robertson-Walker background metric, $m_{P}$
being the Planck mass, and the dot denoting a derivative
with respect to the cosmic time, $t$. The functions $d(q)$ and $f(q)$ are given
by the expressions \cite{Bojowald:2004ax,Bojowald:2003uh}
\begin{align}
	\label{6}
	d(q) =&\left(\frac{3}{2l}\right)^{\frac{3}{2-2l}}q^{3/2}\left\{\frac{1}{2+l}\left[
(q+1)^{l+2}-|q-1|^{l+2}\right]\right.  \nonumber\\
	&\left.-\frac{q}{1+l}\left[(q+1)^{l+1} - \textrm{sign}(q-1)|q-1|^{l+1}\right]\right\}^{\frac{3}{2-2l}}
\end{align}
and
\begin{align}
	\label{7}
	f(q) & = \frac{1}{3}\frac{d }{d \ln (a)}\ln (d) = \nonumber\\
	&=\frac{1}{l-1}\bigg\{(l^{2}-1)\left[(q+1)^{l+2}-|q-1|^{l+2}\right] \bigg.  \nonumber\\
	 - &(2l-1)(l+2)q\left[(q+1)^{l+1} - \textrm{sign}(q-1)|q-1|^{l+1}\right]  \nonumber \\
	\bigg. + &(l+1)(l+2)q^{2}\left[(q+1)^{l}-|q-1|^{l}\right]\bigg\}  \nonumber \\
	\times &\bigg\{(l+1)\left[(q+1)^{l+2}-|q-1|^{l+2}\right] \bigg.  \nonumber \\
	\bigg. - &(l+2)q\left[(q+1)^{l+1} - \textrm{sign}(q-1)|q-1|^{l+1}\right]\bigg\}^{-1}, 
\end{align}
with the definitions $q=(a/a_{*})^{2},$ $a_{*}=(\gamma j/3)^{1/2}l_{P}$, while the
value of the Barbero-Immirzi parameter, $\gamma =0.2375$, is obtained from
black-hole entropy considerations \cite{Meissner:2004ju} (other values can be found in the
literature). The parameters $l$ and $j$ are the so-called ambiguity
parameters and $l_{P}$ is the Planck length. Throughout our paper we use $j=100$ and $l=3/4$. This value of $l$ appears naturally when we derive the
Hamiltonian operator $H_{\phi }$ for the scalar field \cite{Thiemann:1997rt}, while the value of $j$ is set so that the slow-roll inflation lasts for at least 60 $e$-folds \cite{Tsujikawa:2003vr}. We use the
natural system of units with $\rlap{\protect\rule[1.1ex]{.325em}{.1ex}}h=c=1$, and $m_{P}=G^{-1/2}=1.22\times 10^{19}$GeV.

To numerically integrate the equations above, we need to fix the values of
the parameters defining the model and give the initial conditions. We use
the following values: $m_{\phi }=10^{-7}m_{P}$, $a_{i}=\sqrt{\gamma }l_{P}$
, $\dot{\phi _{i}}=0.6\times10^{-6}m_{P}^2$. The value of $\phi _{i}$ is then
obtained by satisfying the uncertainty principle \cite{Tsujikawa:2003vr}
\begin{equation}
	\label{8}
	\left|\phi _{i}\dot{\phi }_{i}\right|\geq \frac{10^{3}}{j^{3/2}}\left(\frac{a_{i}}{a_{*}}\right)^{12}m_{P}^{3}.  
\end{equation}
The value for $\dot{a_{i}}$ is taken as the positive root of equation \eqref{2}, given that the universe is expanding, and this equation is also used to check the accuracy of our integration.

\begin{figure*}[ht]
	\centering
	\subfloat[]{\includegraphics[width=.32\textwidth]{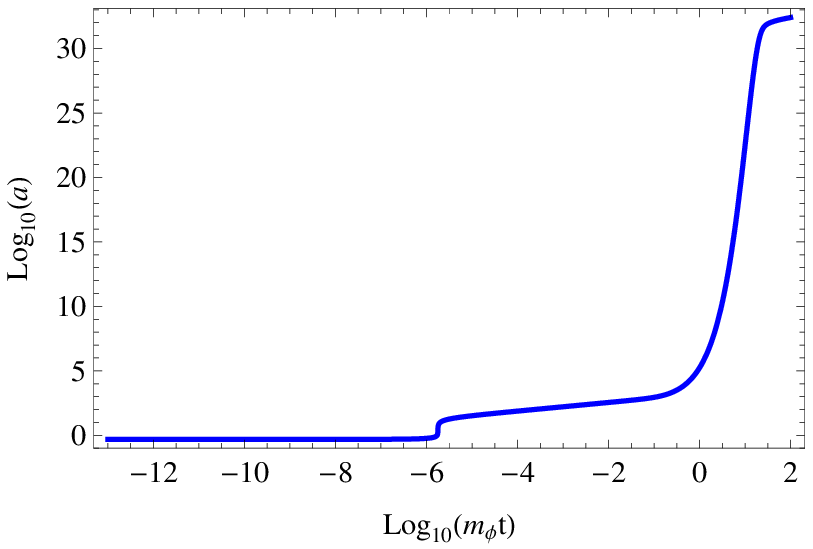}}
	\hfill
	\subfloat[]{\includegraphics[width=.32\textwidth]{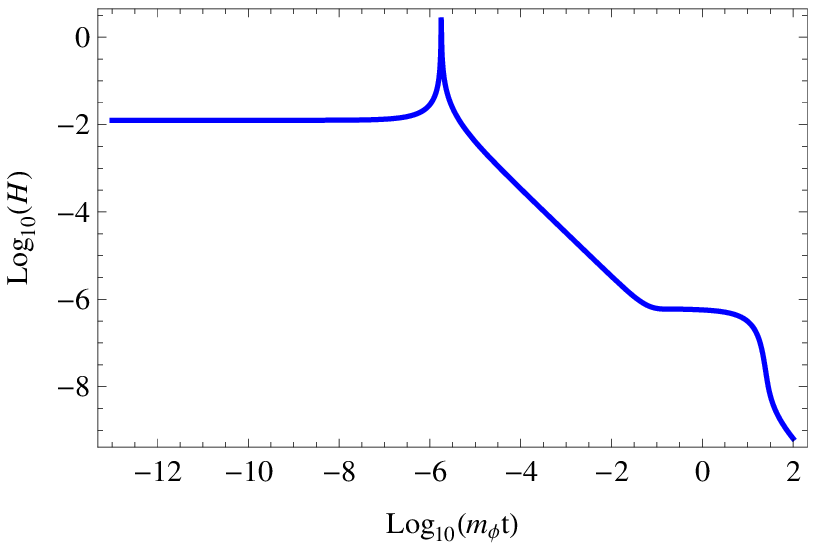}}
	\hfill
	\subfloat[]{\includegraphics[width=.32\textwidth]{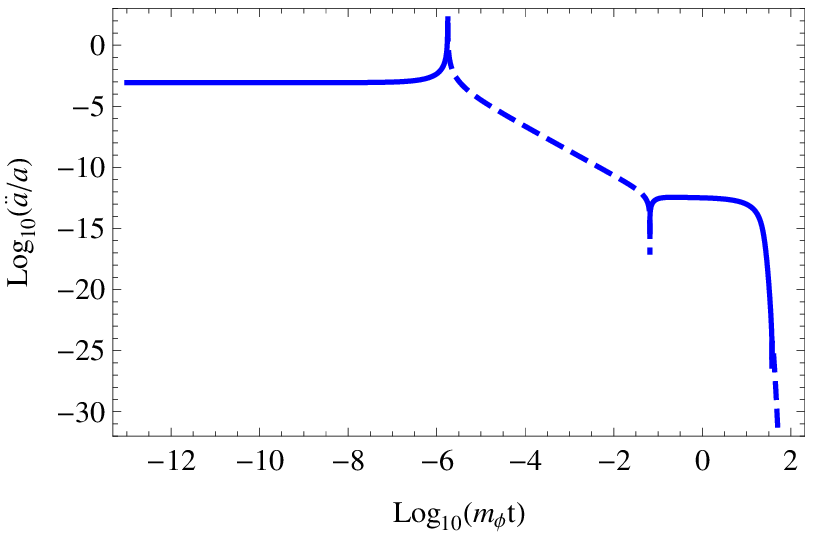}}\\
	\caption[]{\label{fig1}	This figure shows: (a) The scale factor (plotted in units of $m_P^{-1}$) versus the cosmic time. Initially the scale factor is almost constant until it enters the period of super inflation driven by the LQC corrections and, later, the period of classical inflation. (b) the Hubble parameter (plotted in units of $m_P$) versus the cosmic time. The peak on the graphic of the Hubble parameter occurs when $a\approx a_*$. (c) The absolute value of the quotient $\ddot{a}/a$ (plotted in units of $m_P^2$) versus the cosmic time. The continuous line corresponds to the accelerating periods while the dashed line indicates the decelerating periods.}
\end{figure*}

\begin{figure}[h]
	\centering
	\includegraphics[width=.48\textwidth]{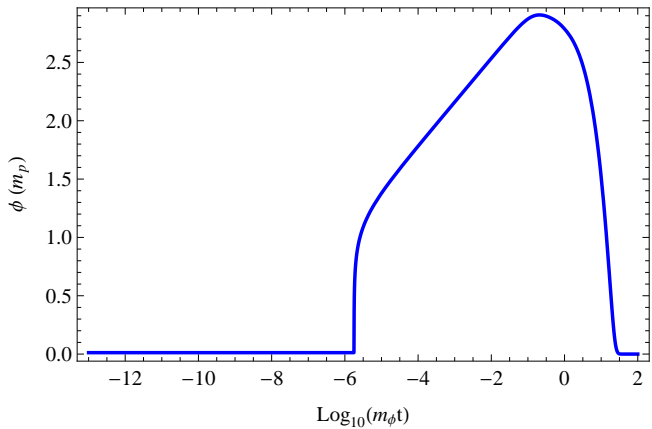}
	\caption[]{\label{fig2} This figure shows the evolution of the scalar field $\phi$ versus the cosmic time. The value of $\phi$ increases abruptly at $a\approx a_*$ and keeps increasing during the deceleration period that ensues until it reaches its maximum value, $\phi_{\textrm{max}}\approx 3.m_P$. Afterwards, the scalar field starts to decrease and enters the classical inflationary era.}
\end{figure}

After a short period of time $d(q)\rightarrow1$ and $f(q)\rightarrow0$ and we enter the classical period, with Eqs.~\eqref{2}, \eqref{3} and \eqref{4} converging to the results of General Relativity. During this period, the scalar field increases from a very small number to $\phi\approx3m_{P}$, enough for a 60 $e$-fold expansion, at which point it begins to decrease, giving way to the standard slow-roll inflation, and oscillate around the minimum of the potential. It is around this time that we switch on the dissipative coefficient $\Gamma_{\phi }$ that governs the energy transfer from the scalar field to a radiation fluid and the reheating of the universe. Therefore, Eqs.~\eqref{2}, \eqref{3} and \eqref{4} are replaced by
\begin{equation}
	\label{5_1}
	\left(\frac{\dot{a}}{a}\right)^2=\frac{8\pi }{3m_{P}^{2}}\left[\frac{\dot{\phi} ^{2}}{2}+V(\phi )\right],  
\end{equation}
\begin{equation}
	\label{5_2}
	\frac{\ddot{a}}{a}=\frac{8\pi }{3m_{P}^{2}}\left[V(\phi)-\dot{\phi} ^{2}\right],
\end{equation}
\begin{equation}
	\label{5_3}
	\ddot{\phi }=-3\frac{\dot{a}}{a}\dot{\phi } - \frac{\partial V}{\partial \phi } - \Gamma _{\phi }\dot{\phi },  
\end{equation}
\begin{equation}
	\label{5_4}
	\dot{\rho _{r}}=-4\frac{\dot{a}}{a}\rho _{r}+\Gamma
_{\phi }\dot{\phi }^{2},  
\end{equation}
with $\rho _{r}$ being the energy of the radiation field, and the dissipative coefficient taking the value $\Gamma _{\phi}=10^{-7}m_{P}$. The evolution of the scale factor $a(t)$, the Hubble parameter $H(t)$, and the quotient $\ddot{a}/a(t)$, until the end of inflation is shown in Fig. \ref{fig1}. The universe goes through an initial super-inflation phase, followed by deceleration, and finally the standard slow-roll inflation that ends at the reheating. The evolution of the scalar field $\phi(t)$ is shown in Fig. \ref{fig2}.

The universe then enters the radiation dominated era. From this point onwards, the equations for the evolution of the universe are given by the $\Lambda$CDM model with a radiation component
\begin{align}
	\label{9}
	\left(\frac{\dot{a}}{a}\right)^{2}=H_0^2\left[\Omega _{r,0}\left(\frac{a_{0}}{a}\right)^{4}+\Omega_{m,0}\left(\frac{a_{0}}{a}\right)^{3}+\Omega _{de,0}\right],  \\
	\label{10}
	\frac{\ddot{a}}{a} = -H_0^2 \left[\Omega_{r,0}\left(\frac{a_{0}}{a}\right)^{4}+\frac{\Omega_{m,0}}{2}\left(\frac{a_{0}}{a}\right)^{3} - \Omega _{de,0}\right].
\end{align}
Here, $\Omega_r$, $\Omega_m$ and $\Omega_{de}$ are the relative densities of radiation, cold matter plus baryonic matter, and dark-energy, respectively. The index 0 indicates the value of a given quantity at the present time. We have set $a_{0}=1$ while imposing that the transition from the reheating to the $\Lambda$CDM model is such that the derivatives $\dot{a}$ and $\ddot{a}$ are continuous. We assign the value $\Omega _{r,0}=0.5\times10^{-4}$, and take the values of the other parameters from the results of the Planck mission \cite{Ade:2013zuv}: $\Omega _{m,0}=0.315$, $\Omega _{de,0}=0.685$, and $H_{0}=67.3$ km/s/Mpc.

These and the equations for the GWs will be numerically
integrated using a fourth-order Runge-Kutta method with variable step.


%
%
\section{\label{sec3}Today's Energy-Spectrum of the Gravitational Waves}

To calculate the energy-spectrum of the cosmological GWs,
generated during the evolution of the universe, we shall be using the method
of continuous Bogoliubov coefficients, first introduced in \cite{Parker:1969au}.
We begin with the wave-equation satisfied by the tensor modes (cf.\cite{Bojowald:2004ax,Bojowald:2003uh} and \cite{Mielczarek:2007zy,Mielczarek:2007wc}),
\begin{equation}
	\label{11}
	\ddot{h} + \left(3H-\frac{\dot{d}}{d}\right)\dot{h}-d\frac{\nabla
^{2}h}{a^{2}}=0,  
\end{equation}
and define the new variable $\mu =ah$; using, for the moment, conformal time 
$d\tau =a^{-1}dt$, we find ($(^{\prime })=d/d\tau $)
\begin{equation}
	\label{12}
	\mu ^{\prime \prime } - \frac{d^{\prime }}{d}\mu ^{\prime }+\left[k^{2}d-\frac{a^{\prime \prime }}{a}+\frac{a^{\prime }}{a}\frac{d^{\prime }}{d}\right]\mu =0, 
\end{equation}
where $d(q)$ is given in Eq.~\eqref{6} above, with $q$ a function of time, and $k=a\omega $, with $\omega$ corresponding to the angular frequency of the GWs.

We now generalize Parker's procedure, introducing the variable
\begin{equation}
	\label{13}
	\mu _{0} = \frac{\stackrel{-}{\mu }}{\sqrt{k}}\sqrt{d(\tau )}\exp
\left(-i\int^{\tau}_{\tau_0}kd\tau ^{\prime }\right),
\end{equation}
where $\bar\mu$ is an arbitrary constant.
Deriving this expression, we can see that $\mu _{0}$ obeys the equation
\begin{equation}
	\label{14}
	\mu _{0}'' - \frac{d^{\prime }}{d}\mu _{0}^{\prime } + \left[k^{2}+\frac{3}{4}\left(\frac{d^{\prime }}{d}\right)^{2}-\frac{1}{2}\frac{d^{\prime \prime }}{d}\right]\mu _{0}=0. 
\end{equation}
It is not difficult to rewrite Eq.~\eqref{12} with the same l.h.s. as in Eq.~\eqref{14}:
\begin{equation}
	\label{15}
	\mu'' -\frac{d^{\prime }}{d}\mu ^{\prime }+ \left[k^{2}+\frac{3}{4}\left(\frac{d^{\prime }}{d}\right)^{2}-\frac{1}{2}\frac{d^{\prime \prime }}{d}\right]\mu =2kS\mu ,  
\end{equation}
with the following espression for $2kS$:
\begin{align}
	\label{16}
	2kS(\tau ) =&\left[k^{2}\left(1-d\right) - \frac{d^{\prime }}{d}\left(\frac{a^{\prime }}{a}-\frac{3}{4}\frac{d^{\prime }}{d}\right)\right. \nonumber\\
	&~~~~~~~~~~~~~~~~~~~~~~~~\left.- \left(\frac{1}{2}\frac{d^{\prime \prime }}{d}-\frac{a^{\prime \prime }}{a}\right)\right],
\end{align}
being the same expression that appears in Eq.~(31) of Ref.\cite{Mielczarek:2007wc}. For
large volumes, $d\rightarrow 1$, $d^{\prime }\rightarrow 0,$ $2kS\rightarrow
a''/a$ and equation \eqref{15} becomes the well-known result of General Relativity
\begin{equation}
	\label{17}
	\mu ^{\prime \prime } + \left(k^{2}-\frac{a''}{a}\right)\mu =0.
\end{equation}

Having reached this point we may now compare Eq.~\eqref{15} with Eq.~(9b) in \cite{Henriques:1993km} and check that they are formally the same except for the more
complicated expression for $2kS$, which in that paper is simply $a''/a$. Following the formulation developed in \cite{Henriques:1993km}, we again obtain
\begin{align}
	\label{mu_alpha_beta}
	\mu = &\frac{\bar\mu}{\sqrt{k}}\sqrt{d(\tau)} \left[ \alpha(\tau) \exp
\left(-i\int^{\tau}_{\tau_0}kd\tau ^{\prime}\right) \right. \nonumber\\
	&~~~~~~~~~~~~~~~~~~~~~~~~\left. + \beta(\tau) \exp
\left(i\int^{\tau}_{\tau_0}kd\tau ^{\prime }\right)\right],
\end{align}
with the Bogoliubov coefficients $\alpha$ and $\beta$ satisfying the relation
\begin{equation}
	\label{24}
	|\alpha |^{2}-|\beta |^{2}=1.  
\end{equation}
From Eq.~\eqref{mu_alpha_beta} we arrive at the differential equations for $\alpha $ and $\beta $:
\begin{align}
	\label{18}
	\alpha ^{\prime }=\frac{i}{2k}\left[\alpha (\tau )+\beta (\tau )e^{2ik(\tau -\tau
_{0})}\right]2kS
\end{align}
\begin{align}
	\label{19}
	\beta ^{\prime }=-\frac{i}{2k}\left[\alpha (\tau )e^{-2ik(\tau -\tau _{0})}+\beta
(\tau)\right]2kS,  
\end{align}
where $2kS$ is given by Eq.~\eqref{16} and where, so far, $\tau _{0}$ is an arbitrary constant. Introducing now the
complex functions $X(k,\tau )$ and $Y(k,\tau )$, through the definitions
\begin{equation}
	 \label{20}
	 X=\alpha e^{-ik(\tau -\tau _{0})}+\beta e^{ik(\tau -\tau _{0})} 
\end{equation}
and
\begin{equation}
	\label{21}
	Y=\alpha e^{-ik(\tau -\tau _{0})}-\beta e^{ik(\tau -\tau _{0})},  
\end{equation}
Eqs.~\eqref{18} and \eqref{19} are replaced by the equations
\begin{equation}
	\label{22}
	X^{\prime \prime } + \left[k^{2} - 2kS(\tau)\right]X=0  
\end{equation}
\begin{equation}
	\label{23}
	Y=\frac{i}{k}X^{\prime }.  
\end{equation}
Notice that at the end of the semi-classical period, as $2kS\rightarrow a''/a$, Eq.~\eqref{22} converges to the result of General Relativity \cite{Sa:2011rm,BouhmadiLopez:2009hv,BouhmadiLopez:2012qp,BouhmadiLopez:2012by}.

We may check that $d^{\prime \prime }(q)$ is singular at the point $q=1$;
this makes it convenient to introduce a new complex variable $Z(\tau )$
\begin{equation}
	\label{25}
	Z=\sqrt{d}X,
\end{equation}
which eliminates the terms with $d''(q)$ in Eq.~\eqref{22}. That differential equation now translates into
\begin{align}
	\label{25a}
	Z'' - \frac{d'}{d}Z' + \left[k^2d + \frac{d'}{d}\frac{a'}{a} - \frac{a''}{a}\right]Z = 0,
\end{align}
and is now suitable for the numerical integration that ensues. Since this integration is done in terms of the cosmological time, $t$, we rewrite Eq.~\eqref{25a} as ($(\dot{\text{ }})\equiv d/dt$)
\begin{equation}
	\label{26}
	\ddot{Z} + \left(\frac{\dot{a}}{a} - \frac{\dot{d}}{d}\right) \dot{Z} + \frac{1}{a^2}\left[k^{2}d + a\dot{a}\frac{\dot{d}}{d} - \left(a\ddot{a}+\dot{a}^{2}\right)\right]Z=0.  
\end{equation}

\begin{figure*}[ht]
	\centering
	\subfloat[]{\includegraphics[width=.48\textwidth]{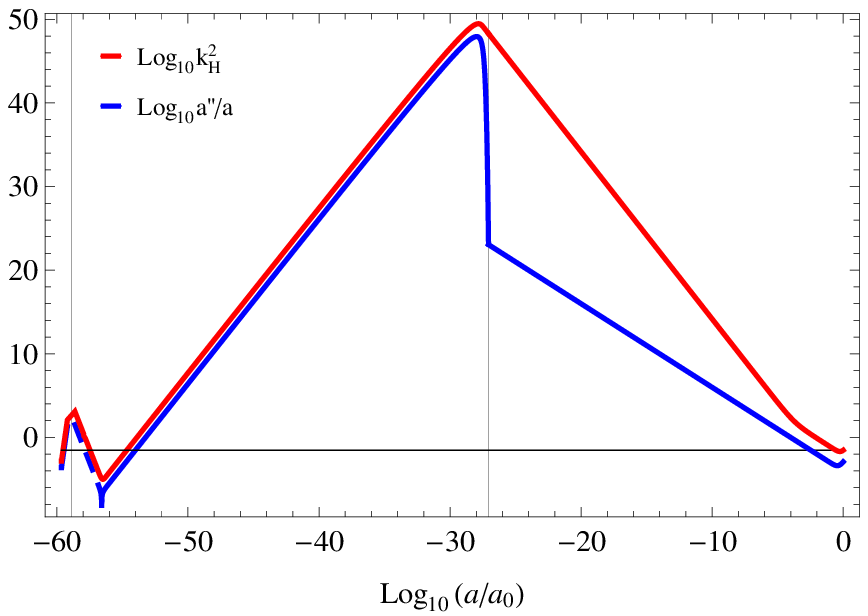}\label{fig4a}}
	\hfill
	\subfloat[]{\includegraphics[width=.48\textwidth]{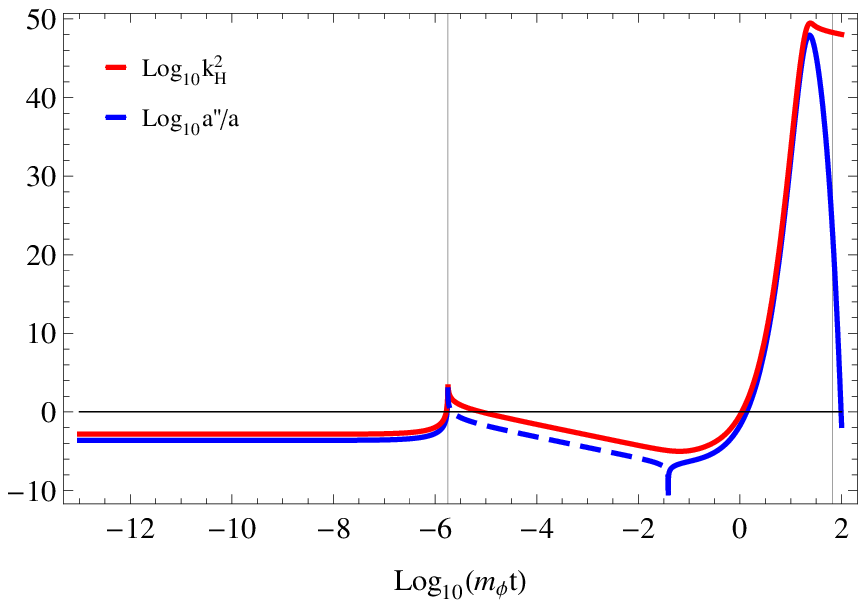}\label{fig4b}}
	\\
	\caption[]{\label{fig4}	This figure shows (a) the evolution of the absolute value of the classical potential $a''/a$ (lower blue line) and the co-moving wave-number $k_H^2=(2\pi aH)^2$ (upper red line) until the present time, (b) zoom of the previous figure: corresponding to the semi-classical period and the inflationary era. The functions are plotted versus the scale factor in Fig. (a) and the cosmic time in Fig. (b). The continuous lines indicate positive values while the dashed line indicates negative values. The horizontal line indicates the mode that is entering the Hubble horizon at the present time. The vertical lines in each figure indicate, from left to right, (i) the point when $a=a_*$, after this point the LQC corrections vanish rapidly and the universe enters the classical evolution; and (ii) the moment of transition to the $\Lambda$CDM model with a radiation component.}
\end{figure*}

We next integrate numerically this equation, using the results of Sec.~\ref{sec2} for the evolution of the quantities $a(t)$ and $d(t)$ and their derivatives. The integration is done through the various stages of evolution of the universe, from the semi-classical period, followed by the classical inflation and the reheating. At this point we change variables once more and perform the integration during the radiation-dominated, the matter-dominated, and finally the dark-energy-dominated eras, until the present time, in terms of the scale factor. In Fig.~\ref{fig4} we show the evolution of the co-moving wave-number, $k_H^2=(2\pi aH)^2$, and the classical potential $a''/a$. During the classical regime, the production of gravitons occurs for each mode when $k^2\ll a''/a$, i.e., when the mode is well inside the Hubble horizon as $a''/a$ is roughly of the order of $k_H^2$, see Fig.~\ref{fig4}.

The power-spectrum $P(\omega )$ is given by \cite{Allen:1987bk}
\begin{equation}
	\label{(27)}
	P\left(\omega\right)=\frac{\rlap{\protect\rule[1.1ex]{.325em}{.1ex}}h\omega ^{3}}{\pi ^{2}c^{3}}\left|\beta _{final}\right|^{2},  
\end{equation}
in units erg.s.cm$^{3}$. We shall express our results in terms of the relative logarithmic energy-spectrum of the GWs, defined as
\begin{equation}
	\label{28}
	\Omega _{GW}\left(\omega ,t_{0}\right) = \frac{1}{\rho _{c}}\frac{d\rho _{gw}}{d\ln\omega },  
\end{equation}
where $\rho _{c}$ is the value of the present time critical density and $\rho _{gw}$ is the GW energy density, 
\begin{equation}
	\label{29}
	\rho _{gw}=\int P(\omega )d\omega .  
\end{equation}
The final expression for $\Omega (\omega ,t_{0})$, in terms of present day values, is then \cite{Allen:1987bk}
\begin{equation}
	\label{30}
	\Omega _{GW}=\frac{8\rlap{\protect\rule[1.1ex]{.325em}{.1ex}}hG}{3\pi c^{5}H_{0}^{2}}\omega _{0}^{4}|\beta _{0}|^{2}.  
\end{equation}

The number of gravitons, $\langle n_g(t) \rangle$, at any time $t$ is related to the Bogoliubov coefficient $\beta (t)$, and can be expressed in terms of the complex functions $X(t)$ and $Y(t)$ as
\begin{equation}
	\label{31}
	\langle n_g(t) \rangle = \left|\beta (t)\right|^{2}=\frac{1}{4}\left[X(t)-Y(t)\right]\left[X^{*}(t)-Y^{*}(t)\right],  
\end{equation}
the $*$ denoting complex conjugation. After the integration, $X(t)$ is obtained from $Z(t)$ through \eqref{25} (actually, at the end of integration $d(t)=1$ and $X=Z$) and $Y(t)$ is given in Eq.~\eqref{23}, which, in terms of cosmic
time $t$, becomes
\begin{equation}
	\label{32}
	Y(t)=i\frac{a_{0}\omega _{0}}{a(t)}X(t).  
\end{equation}
For simplicity, we assumed that at the beginning of the integration no
GWs were present, choosing $X(t_{i}),$ $Z(t_{i})$ and $Y(t_{i})$ such that $\beta(t_{i})=0$ and $\alpha(t_i)=1$ \cite{Parker:1969au}.

\begin{figure}[ht]
	\centering
	\includegraphics[width=.48\textwidth]{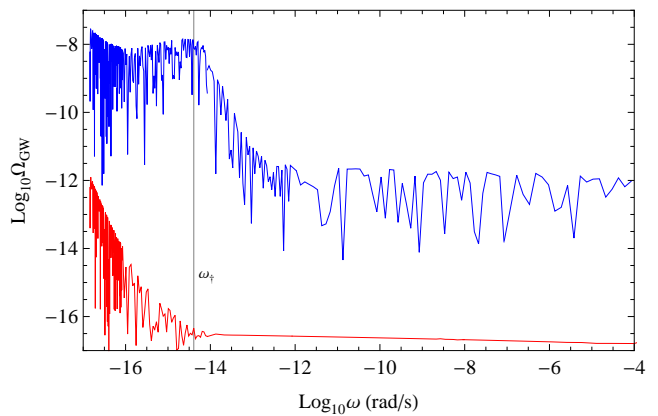}
	\caption[]{\label{fig5}	This figure shows the relative logarithmic energy-spectrum of the cosmological GWs as could be measured today, calculated with (upper blue line) and without (lower red line) the inverse-volume corrections of LQC in the evolution equations of the tensor perturbations. The vertical line indicates the frequency $\omega_\dagger$, which corresponds the peak of the module of the term $a\dot{a}\frac{\dot{d}}{d} - \left(a\ddot{a}+\dot{a}^{2}\right)$ in Eq.~\eqref{26}}
\end{figure}

The results for the energy-spectrum are shown in Fig.~\ref{fig5}. In this figure we compare our results with those obtained with exactly the same background evolution, but without inserting, in the GW equations, the inverse-volume corrections \cite{Sa:2011rm}. We see that, while both spectra show a rise of the energy density of the GWs with respect to General Relativity, there are some important differences between them.

First, when the LQC corrections are introduced only at the background level, the imprints of those corrections appear only on the low-frequency end of the spectrum \cite{Afonso:2010fa,Sa:2011rm}; in the present case large oscillations appear on the spectrum up to frequencies of the order of $10^{-15}$rad/s. This is due to the fact that, if the tensorial perturbations are treated like in standard General Relativity, i.e. we set $2kS=a''/a$ in Eq.~\eqref{22}, the function $2kS$ acts as a potential for the production of gravitons, which are created whenever the condition $k^2\ll a''/a$ is satisfied\footnote{In the regime $k^2\gg |2kS|$ or when $2kS$ is constant, Eq.~\eqref{22} admits oscillatory sinusoidal solutions and so $|\beta|^2$ remains constant.}. We can obtain an estimate of the maximum frequency for which the oscillations appear, by calculating the frequency $\omega_*$ that corresponds to the maximum of $a''/a$ during the semi-classical period, see the most leftward peak in Fig.~\ref{fig4b}. The value obtained for $\omega_*$ was
\begin{align}
	\omega_* \approx3.19\times10^{-15} \textrm{ rad/s},
\end{align}
which is in agreement with the results of Fig.~\ref{fig5}. For higher frequencies the spectrum becomes almost flat and horizontal. 

However, when we consider the inverse-volume corrections of LQC at the background and at the perturbative level, we observe a considerable growth of the energy density of the GWs, up to three/four orders of magnitude with respect to the results with the loop corrections only at the background level. This effect does not appear to have a clear cut-off frequency, as it extends to frequencies of the order of $\sim10^{-4}$ rad/s. We can explain this effects in light of the modifications introduced in Eq.~\eqref{22} by the LQC corrections at the perturbative level. Analysing Eq.~\eqref{16}, we find that the specific form of $2kS$ cancels the constant term $k^2$ in Eq.~\eqref{22}, which can now be recast as
\begin{align}
	X'' + \left[k^2d(\tau) - U(\tau)\right]X = 0.
\end{align}
Here, $U(\tau)=\frac{d^{\prime }}{d}\left(\frac{a^{\prime }}{a}-\frac{3}{4}\frac{d^{\prime }}{d}\right) - \left(\frac{1}{2}\frac{d^{\prime \prime }}{d}-\frac{a^{\prime \prime }}{a}\right)$ is independent of $k$ and approaches $a''/a$ at the semi-classical period. As it contains terms with the second derivative of $d(\tau)$, the function $U(\tau)$ is singular at $q\approx1$, which seems to induce a very intense production of gravitons during the intial super-inflationary phase. Furthermore, the fact that the term in $k^2d(\tau)$ is no longer constant means that the higher modes are not ``blind'' to the effects of the LQC corrections at the perturbative level, in contrast with what happens when the loop corrections are considered only at the background level. 

Furthermore, we observe the presence of a local broad maximum in the frequency range $10^{-15}\sim10^{-14}$ rad/s which is absent when the inverse-volume corrections are not included at the perturbative level. Upon inspection we found that the position of the maximum on the spectrum is consistent with the frequency $\omega_\dagger$, corresponding to the maximum value of the term $a\dot{a}\frac{\dot{d}}{d} - \left(a\ddot{a}+\dot{a}^{2}\right)$, see Eq.~\eqref{26}, during the initial super-inflationary phase. The value calculated for $\omega_\dagger$ is 
\begin{align}
	\omega_\dagger \approx 4.06\times 10^{-15}\textrm{ rad/s}.
\end{align}

Another feature of the energy-spectrum of the GWs is the initial slope which appears near the minimum frequency, $\omega_{\textrm{hor}}\approx1.4\times10^{-17}$ rad/s, corresponding to the current horizon. This slope appears from the combination of (i) the loop corrections in the tensorial equations, and (ii) the production of gravitons during the matter-dominated phase. While the first increases the energy density of the GWs by several orders of magnitude, as mentioned above, the second originates an additional raise of the energy density only on the very low frequencies range, on the left of the maximum that occurs at $\omega\approx\omega_\dagger$.

Due to the large amount of time necessary to produce each point of the spectrum, above $10^{-12}$ rad/s, we did not complete the spectrum beyond the frequency $10^{-4}$ rad/s.

%
%

\section{\label{sec4}Conclusions}

In this work we have investigated the energy-spectrum for the gravitational
waves generated within a loop quantum cosmological model. The evolution of
the universe, as here modelled, goes basically through two stages,
first a semi-classical stage with super-inflation, whose equations receive
important corrections coming from LQC, followed by a
classical evolution described by the usual general relativistic equations.
In the semi-classical stage, the corrections we introduced were of the
inverse-volume type, leaving to a future work the study of the influence of
the holonomy corrections. These corrections, particularly to the inflation equation, push up the value of the
scalar field, giving rise, in a natural way, to those values of the order of
Planck mass which are necessary to obtain enough inflation.

By numerically integrating the equations, introduced in Sec.~\ref{sec2} and \ref{sec3},
we were able to calculate the relative logarithmic energy-spectrum $\Omega
_{\textrm{GW}}$ for the GWs, as would be seen today. In fact, we
calculated two spectra, one with the inverse-volume corrections inserted in
all the dynamical equations, including the GW equations, the
other where those corrections were only included in the equations governing
the expansion of the universe, but not in the equations for the
GWs, as was seen before in Ref.~\cite{Sa:2011rm}. The physical processes taking place before the standard slow-roll inflation leave their imprint on the spectrum in the
region of very low frequencies. Considerable differences were observed in
the two situations, demonstrating the importance of the inverse-volume
corrections, for the production of gravitons. First, we have an important
extra production of gravitons, by at least three orders of magnitude and,
second, we observe a local maximum around $\omega_\dagger\approx4\times10^{-15}$ rad/s. This is consistent with a resonance at the frequency corresponding to the peak of the term $a\dot{a}\frac{\dot{d}}{d} - \left(a\ddot{a}+\dot{a}^{2}\right)$ in Eq.~\eqref{26}, which occurs at the end of the initial super-inflation epoch driven by loop effects, i.e. at $q\approx1$. This maximum is absent when the LQC corrections are not included in the GW equations. Finally, for frequencies above $10^{-12}$rad/s, the spectrum, instead of becoming almost flat and horizontal, continues to show important oscillations in a large interval of frequencies, at least up to $10^{-4}$ rad/s.
We would like to highlight once more that our calculations involve only inverse volume corrections, at the background and perturbative levels. Therefore, we have disregarded the holonomy corrections . The later are very important on LQC as they remove the Big Bang singularity through a bounce. If the bounce is located around $a_*$, the regime $a\leq a_*$ is not reached, which is precisely where we have an overproduction of gravitons. Therefore, it could be that the inclusion of the holonomy corrections remove or appease this overproduction.\footnote{We are very grateful to the anonymous referee for this important remark.}
We will tackle this issue on the near future.\\

\section{Acknowledgements}

M.B.L. is supported by the Basque Foundation for Science IKERBASQUE.
This work was supported by the Portuguese Agency “Funda\c{c}\~ao para a Ci\^encia e Tecnologia” through PTDC/FIS/111032/2009 and partially by the Basque government Grant No. IT592-13.

\end{document}